\documentclass[final]{ias2}

\usepackage{graphicx} 
\usepackage{multirow}
\usepackage{array} 
\usepackage{amsmath,amssymb}
\usepackage[ps2pdf]{hyperref}
\usepackage{textcomp}

\newcommand{\tc}{T_c}
\newcommand{\qq}{\bar{Q}Q}

\newcommand{\qqc}{\bar{c}c}
\newcommand{\jpsi}{J/\psi}
\newcommand{\jsub}{{\scriptstyle J/\psi}}
\newcommand{\psib}{\psi^\prime}
\newcommand{\yn}{\Upsilon}
\newcommand{\yna}{\Upsilon(1S)}
\newcommand{\ynb}{\Upsilon(2S)}
\newcommand{\ync}{\Upsilon(3S)}
\newcommand{\beq}{\begin{equation}}
\newcommand{\eeq}{\end{equation}}
\newcommand{\ber}{\begin{eqnarray}}
\newcommand{\eer}{\end{eqnarray}}
\newcommand{\xt}{\vec{x},t}
\newcommand{\al}{\alpha_s}
\newcommand{\xtau}{\vec{x},\tau}

\newcommand{\om}{\omega}
\newcommand{\som}{\rho_{\scriptscriptstyle H}(\omega)}
\newcommand{\default}{\rho_0(\omega)}
\newcommand{\somp}{\rho_{\scriptscriptstyle H}(p_0, \vec{p})}
\newcommand{\somw}{\rho_{\scriptscriptstyle H}(\omega, \vec{p})}
\newcommand{\mq}{m_{\scriptscriptstyle{Q}}}
\begin{document}

\markboth{Quarkonia in heavy ion collisions}{S. Datta}

\title{Quarkonia at finite temperature in relativistic heavy ion collisions}

\author[sin]{Saumen Datta} 
\email{saumen@theory.tifr.res.in}
\address[sin]{Department of Theoretical Physics, Tata Institute of 
Fundamental Research, Homi Bhabha Road, Mumbai - 400005, India.}

\begin{abstract}
The behavior of quarkonia in relativistic heavy ion collisions is reviewed.
After a detailed discussion of the current theoretical understanding 
of quarkonia in a static equilibriated plasma, we discuss 
quarkonia yield from the fireball created in ultrarelativistic 
heavy ion collision experiments. We end with a brief discussion of the 
experimental results and outlook.
\end{abstract}



\maketitle


\section{Introduction}
\label{sec.intro}
The connection between quarkonia and deconfinement began with the remarkable
paper of Matsui and Satz \cite{matsui}. The basic idea is extremely
simple.  At high temperatures, due to Debye screening the binding
between a $\qq$ pair becomes of the Yukawa form, and for sufficiently
high temperatures the $\qq$ meson does not form as the binding becomes 
sufficiently weak.  For $\jpsi$ the temperature was estimated to be
very close to $\tc$, the temperature for transition from a hadronic
state to a deconfined plasma.  Because the $\jpsi$ shows a prominent
peak in the dilepton channel, the disappearance of this peak would be
the indicator of deconfinement. In follow-up papers \cite{kms,digal},
the dissociation temperature of various quarkonia was calculated using
the Debye screened form. It was found that the 1S charmonia would
decay very close to the transition temperature while the 1P, 2S etc
states would decay even earlier.  On the other hand, the $\yn$
was found to survive till much higher temperatures.  Therefore the
quarkonia states were suggested as a thermometer for the plasma.

A suppression of the $\jpsi$ peak was indeed found in the fixed target
158A GeV Pb-Pb collisions \footnote{Throughout this report, we will be
  discussing nucleus-nucleus collisions, with a given energy per
  nucleon. In such caes, the energy will be written as xxxA GeV, where
  xxx GeV is the energy per nucleon and A is the mass number.}  in the
NA50 experiment in CERN \cite{na502}. 
Similar suppression has also been seen in the colliding
machine experiments at 200A GeV Au-Au collision in RHIC, and at 2.76A
TeV Pb-Pb collisions in LHC. 

A more detailed theoretical analysis of the behavior of quarkonia in
quark-gluon plasma (QGP), however, has shown more intricacies than
originally thought. Even in the case of static equilibrium plasma,
theoretically the simplest one to handle, the behavior of quarkonia
seems quite complicated in the temperature regime $1-3 \tc$ which is of
interest to relativistic heavy ion collision
experiments. Experimentally, there is strong evidence that a
deconfined medium has been formed in relativistic heavy ion collision
experiments. While suppression of quarkonia have been a generic
feature in these experiments, the detailed behavior has been more
complicated to understand. Quarkonia remain among the most studied 
observables in such experiments; but probably they provide more an
insight into the nature of the plasma rather than act as a thermometer. 

Here we will review our current understanding of the theory of
quarkonia in deconfined medium. In the next section, we will discuss
in some detail the idealized problem of quarkonia in an equilibriated
plasma at a fixed, not-too-high temperature. In Sec. \ref{sec.hic}
we will discuss attempts to study quarkonia in the fast
expanding fireball that is created in the experiments. Section
\ref{sec.expt} contains a short outline of the main experimental
results, for completeness. The last section contains a summary.

\section{Quarkonia in static equilibrium plasma}
\label{sec.theory}
In this section we will discuss our current theoretical understanding of, 
and challenges in understanding of, the behavior of quarkonia in
deconfined plasma. For definiteness, we will consider the case of a
$\qq$ pair in a definite quantum number channel, put in as probe of
the medium, where $Q$ can be charm or bottom.

As outlined in Sec. \ref{sec.intro}, the first studies of quarkonia in
QGP were built on the reasonably successfull nonrelativistic potential
model approach to quarkonia spectroscopy. Instead of a confining
potential, the Debye screened form of the potential, \beq V(r) \ =
\ -\frac{4}{3} \frac{\alpha_s}{r} e^{- m_D r}
\label{eq.debye}
\eeq was used. Here $\alpha_s$ is the strong coupling constant and
$m_D$, the debye mass in QGP. The aim was to calculate a dissociation
temperature for the different quarkonia by solving the Schr\"odinger
equation with $V(r)$. The perturbative form of the potential was later
substituted by the free energy of a static $\qq$ pair, calculated
from lattice.

Use of Eq. (\ref{eq.debye}) in this way, however, is not based on
strong theoretical footing. Recent attempts to understand the behavior
of quarkonia in-medium have started with a rephrasing of the question:
e.g., ``what happens to the $\jpsi$ peak in the dilepton channel if a
plasma is formed?'' is best understood by looking at a quantity that
directly looks at the dilepton channel. $\jpsi$ connects to the dilepton channel
by the point vector current $V_i(x)=\bar{c}(x) \gamma_i c(x)$, and
the suitable correlator of this current controls the dilepton
rate. The dilepton rate can be directly connected to the spectral function, 
which is the fourier transform of the correlator, \cite{lebellac} \\
\beq
\somp \ = \ \int dt \, \int d^3x \, e^{i p_0 t -
  i \vec{p} \cdot \vec{x}} \: \langle [J_H(\xt) , J_H(\vec{0}, 0)] \rangle
\label{eq.sigma}
\eeq
where $J_H(\xt) = \bar{Q}(\xt) \gamma_H Q(\xt)$ is the suitable hadronic
point current, and the angular bracket indicates thermal averaging. 
$\somp$ is called the spectral
function and is proportional to the dilepton rate.

A mode expansion of $\somp$ is instructive. Inserting a complete set
of states, one can write $\somp$ as \\
\beq
\somp \ = \ \frac{1}{Z} \ \sum_{n,m} \left( e^{-k^0_n/T}-e^{-k^0_m/T} \right) 
\ \lvert 
\langle n \lvert J_H(0) \rvert m \rangle \rvert^2 \ 
\delta^4(p^\mu - k^\mu_m + k^\mu_n)
\label{eq.mode} \eeq 
where the sum over states includes both discrete and continuous
states, and $k^\mu_n$ is the four-momenta of the state $\vert
n \rangle$. 

In case of a free scalar particle of mass $M$, the expression above
leads to a spectral function
\beq
\somp \vert_{\rm free} \ = \ \epsilon(p_0) \: \delta(p^2 - M^2).
\label{eq.free} \eeq
As $T \to 0$, a stable meson contributes a similar term to the
spectral function in QCD (with a multiplicative factor $\lvert \langle
0 \lvert J_H \rvert M \rangle \rvert^2$). When the state is unstable, the
delta function gets smeared into a smooth peak, whose width
reflects the decay width of the particle. For a particle like the
$\jpsi$ with a narrow decay width, one gets an almost-$\delta$ function peak,
which shows up in the dilepton cross section. At finite temperatures, as
Eq. (\ref{eq.mode}) shows, we will have a more complicated expression;
the question of interest is whether the peak structures corresponding
to various quarkonia survive at a given temperature.

\subsection{Spectral function using lattice QCD}
\label{sec.lattice}

QCD, the theory of strong interactions, cannot be directly defined on the 
continuum (like other quantum field theories), and needs to be regularized. 
Regularization using a space-time lattice has proved invaluable for 
studies of the nonperturbative regime of QCD, as one can use numerical
Monte Carlo techniques. Much of our current knowledge of strongly
interacting matter at moderately high temperatures (that are of
interest to the ultrarelativistic heavy ion collision experiments), in
particular the transition temperature, equation of state, nature of
the transition, etc, comes from lattice QCD \cite{eos_review}. 

Since we are interested in the spectral function, $\rho$, at
temperatures $\lesssim 3 \tc$, where perturbation theory may not work
very well, it would be ideal to calculate the spectral function,
$\rho(\omega=p_o, \vec{p})$ using lattice QCD. The catch is that lattice
QCD is defined in Euclidean time, and the thermal correlators one
can calculate numerically are the Matsubara correlators 
\beq
C_H(\xtau) \ = \ \langle J_H(\xtau)  J_H(\vec{0}, 0) \rangle
\label{eq.matsubara} 
\eeq
where $\tau \in [0, \beta=1/T)$ is defined in the Euclidean time
direction. In order to get the real time correlators of
Eq. (\ref{eq.sigma}) one needs to use an analytic continuation in
time, $\tau \to -i t$. This leads to the following integral equation
connecting $\somw$ and $C_H(\xtau)$: 
\begin{subequations}
\ber
C_H(\tau, \vec{p}) = \int d^3 x \: e^{-i \vec{p} \cdot \vec{x}} \:
C_H(\xtau) 
& = & \int d\om \: \somw \ K(\omega, \tau)
\label{eq.int} \\
K(\omega, \tau) &=& \frac{\cosh \om(\tau - 1/2T)}{\sinh \om/2T} .
\label{eq.kernel} \eer
\label{eq.connect}
\end{subequations}
For $T=0$, Eq. (\ref{eq.int}) simplifies to a Laplace transform. In the rest
of this section, we will mostly consider correlators projected to
$\vec{p}=0$, and denote them simply as $C_H(\tau)$, omitting the
$\vec{p}$ argument. The corresponding spectral function $\rho_H(\om,
\vec{p}=0)$ will be written as $\som$.

The first lattice studies of $\jpsi$ and other charmonia states are
about a decade old \cite{prd,matsufuru,asakawa}. They all used the
``quenched approximation'', i.e., the plasma was purely
gluonic. $\mathcal{O}(10)$ (12-32) data points were used in the $\tau$
direction, and the spectral function was estimated using
Eq. (\ref{eq.int}). An examination of Eq. (\ref{eq.int}) immediately
shows the difficulty of the extraction of $\som$ from $C(\tau)$: the
inverse Laplace transform is a very nontrivial problem numerically,
made even more difficult by the dual facts of the small range of
$\tau$ at high temperatures and the $\mathcal{O}(10)$ data
points. Note that because of the periodicity of the kernel in
Eq. (\ref{eq.kernel}), one has independent information about Matsubara
correlation only for $\tau \in [0, \beta/2)$. Clearly, a direct
inversion of Eq. (\ref{eq.connect}) is not possible. The studies used
the maximum entropy method (MEM) \cite{mem}, where Bayesian theory is
used to provide information about the solution $\som$ when it is not
constrained by the data. If one treated the extraction of $\som$ from
$C(\tau)$ as, e.g., a simple $\chi^2$ minimization problem, one would
have many flat directions in the parameter space. In maximum entropy
method, one stabilizes the analysis by recasting it as a maximization
of the combination 
\begin{subequations}
\begin{eqnarray}
\mathcal{L} & = & - \frac{1}{2} \: \chi^2 \: + \: \alpha \, S ,
\label{eq.mem} \\
S & = & \sum_i \Delta \om_i \: \left( 
\rho_H(\omega_i) - \rho_0(\omega_i) - \rho_H(\omega_i) \log \frac{\rho_H(\omega_i)}{\rho_0(\omega_i)}
\right) .
\label{eq.entropy} 
\end{eqnarray}
\label{eq.bayes} 
\end{subequations} 
Here $\default$ is the default solution provided as an input to
the analysis; in the absence of data, Eq. (\ref{eq.bayes}) implies
that $\som = \default$. Eq. (\ref{eq.entropy}) has the form of entropy
in information theory, giving the method its name. Given a set of
$C(\tau)$  the maximization of $\mathcal{L}$, Eq.(\ref{eq.mem}), has a unique
solution \cite{mem}. Note that this by itself does not guard against
a solution unstable against noise. As pointed out by Bryan \cite{bryan}, 
parametrized suitably, the solution space can be restricted to the 
space spanned by singular directions \cite{press} of the kernel in 
Eq. (\ref{eq.int}), whose dimensionality is no more than the number of
data points.

\begin{figure}[ht]
\begin{center}
\includegraphics[width=0.42\textwidth,height=0.35\textwidth]{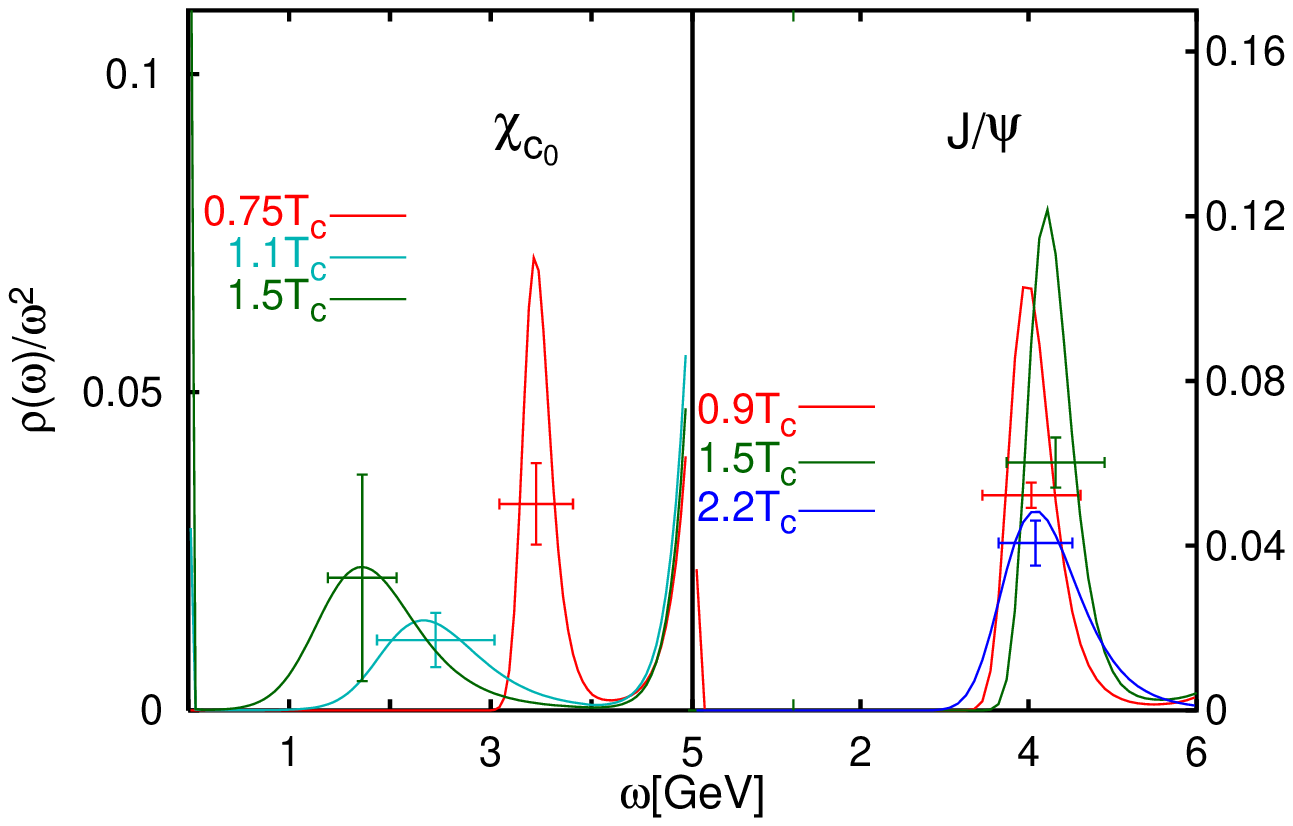}
\includegraphics[width=0.4\textwidth,height=0.32\textwidth]{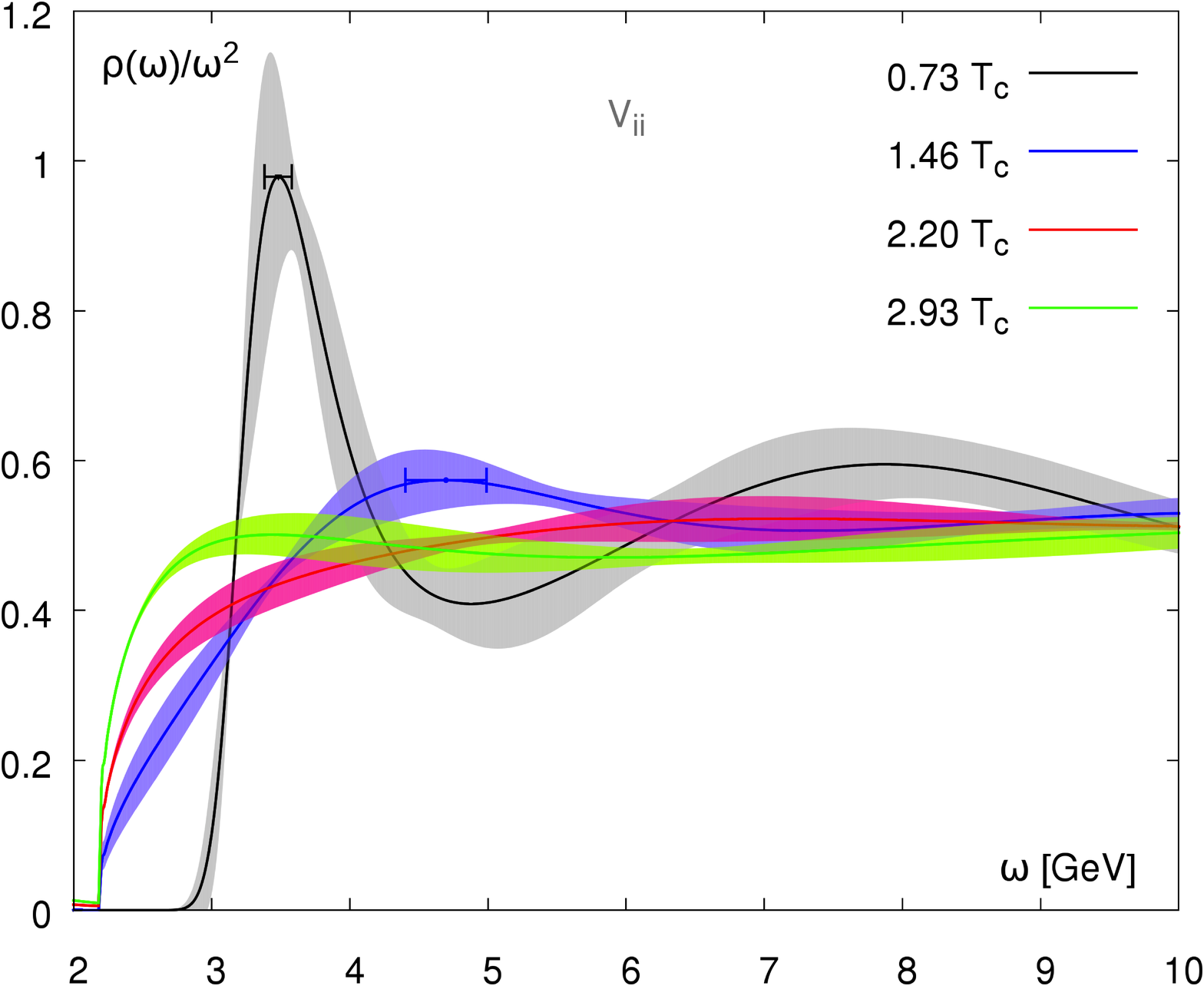}
\caption{(Left) Spectral functions of the $\bar{c}c$ and $\bar{c} \gamma_i c$ 
operators from Ref. \cite{prd}, showing the modification of the
$\chi_{c_0}$ and survival of the $\jpsi$ peak after deconfinement. The vertical 
error bar indicates the error for the reconstructed spectral function, 
averaged over the $\omega$ range indicated by the horizontal band \cite{mem}.
(Right) Spectral function of the $\bar{c} \gamma_i c$ operator
(\cite{ding}, \textcopyright American Physical Society) showing
serious modification of the $\jpsi$ peak
already below 1.5 $\tc$. The band is the error estimate from a 
simple jackknife analysis.}
\label{fig.jpsi}
\end{center}
\end{figure}

While the early studies, Ref. \cite{prd,asakawa,matsufuru}, differed
in some details, they all found that the spectral function of
$\jpsi$ was not very sensitive to the phase transition: the changes in
$\som$ were small as one crossed $\tc$. A clear peak structure was
found even at 1.5 $\tc$. In addition, the dissolution of the peak was
found to be gradual rather than abrupt \cite{prd, matsufuru}, with a
broadening and weakening of the peak as one went to higher
temperatures. On the other hand, the 1P states ($\chi_c$) were seen to
change much more abruptly across the transition. The spectral
functions calculated in Ref. \cite{prd} for the scalar, $\bar{c} c$,
and vector, $\bar{c} \gamma_i c$, operators are shown in the left
panel of Fig. \ref{fig.jpsi}. The correlators $c(\tau)$ show very
little change in the vector channel as one crosses $\tc$, even upto
1.5 $\tc$; this resulted in an extracted spectral function that showed
a strong $\jpsi$ peak even at 1.5 $\tc$. On the other hand, $C(\tau)$
in the scalar channel showed serious modification on crossing $\tc$,
and major weakening of the $\chi_{c_0}$ peak was seen already at 1.1 $\tc$.
Several follow-up studies reached qualitatively similar conclusions
\cite{veletsky}. Also a dynamical study with 2-flavor QCD (but with a
somewhat heavy pion) found very similar results, when temperatures are
expressed in units of $\tc$ \cite{skullerud}. A very recent dynamical study,
again with a somewhat heavy pion, has also found very little change in the
1S state peaks upto temperatures $\sim 1.4 \tc$ \cite{fodor}.

On the other hand, the systematics of the inversion of
Eq. (\ref{eq.connect}) is large, and probably the extraction of $\som$
was less reliable than what the convergence of the different results
suggested.  In particular, it was pointed out later that a large part
of the change in the 1P channels is due to the diffusion peaks in
these channels \cite{umeda}. The free spectral function in the vector,
axial vector, and scalar channels have a contribution
\beq
\som \xrightarrow{\om \to 0} 2 \pi \: \chi_H(T) \: \om \delta(\om)
\label{eq.low} \eeq
which contribute an additive constant to $C(\tau)$. In the interacting
theory the $\delta$ function becomes a smooth peak, leading to a
near-constant term in the correlator. It has been shown
\cite{umeda,peter} that much of the change in the 1P channel
correlators comes from this low-$\omega$ contribution.  It was also
pointed out that even at small temperatures $T \sim 0$, when one
restricts oneself to a small range in Euclidean time, it is difficult
to isolate the peak structure from Euclidean correlator data
\cite{veletsky}. In a series of papers Mocsy and Petreczky
\cite{mocsy} have shown that the lattice data of $C(\tau)$ is also
consistent with a large change in the peak structure at relatively
smaller temperatures.

A recent study, similar in approach to Refs. \cite{prd,asakawa} but 
using finer lattices\cite{ding}, has
found that even in the vector and pseudoscalar channels, the 
peak structure is considerably softened already at 1.5 $\tc$. Results 
from this study are also shown in the right panel of
Fig. \ref{fig.jpsi}. 
It is to be understood that the analysis is
similar to the earlier lattice studies, and can suffer from similar
systematic effects as those.  While it is certainly reasonable to hope
that lattice QCD will be able to provide the spectral function with
much better systematics in the future, it would be important to
incorporate new ideas into the calculation.

\begin{figure}[ht]
\begin{center}
\includegraphics[width=0.55\textwidth,height=0.4\textwidth]{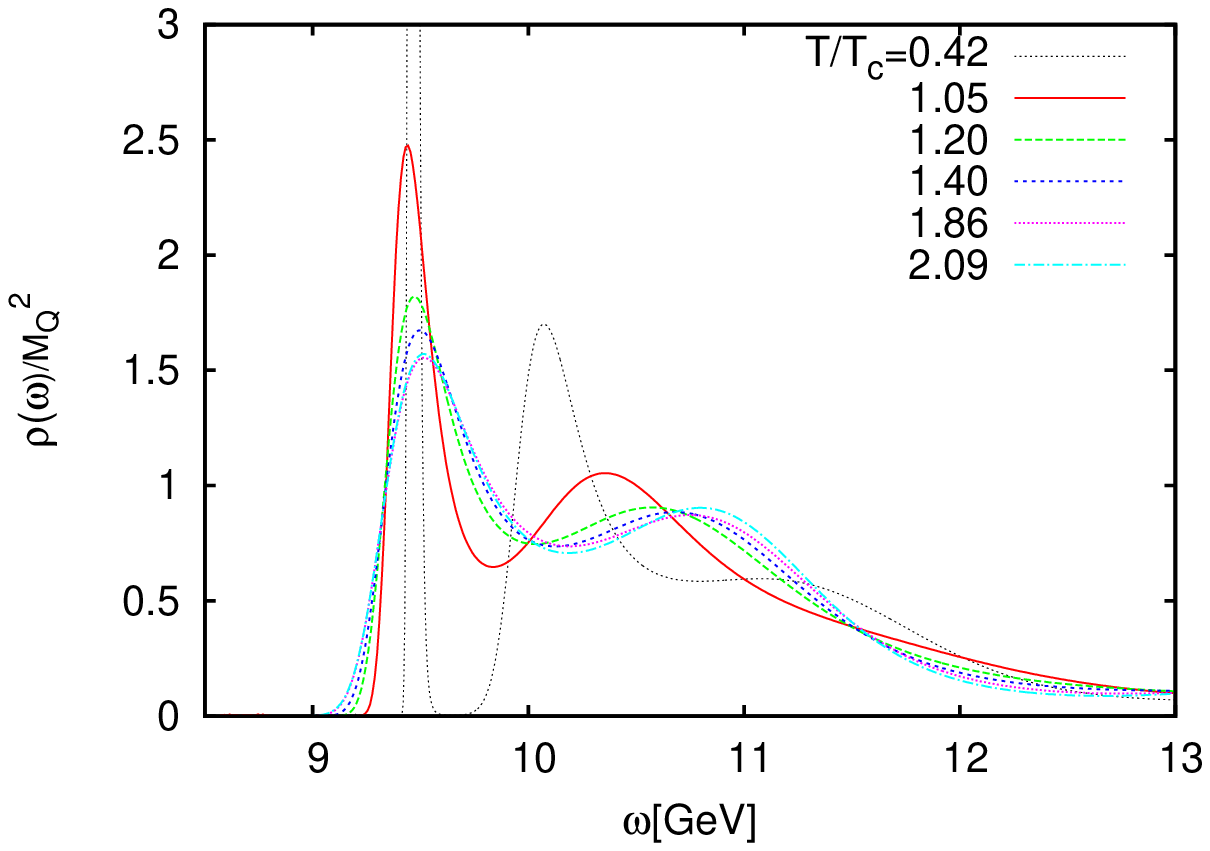}
\includegraphics[width=0.35\textwidth,height=0.4\textwidth]{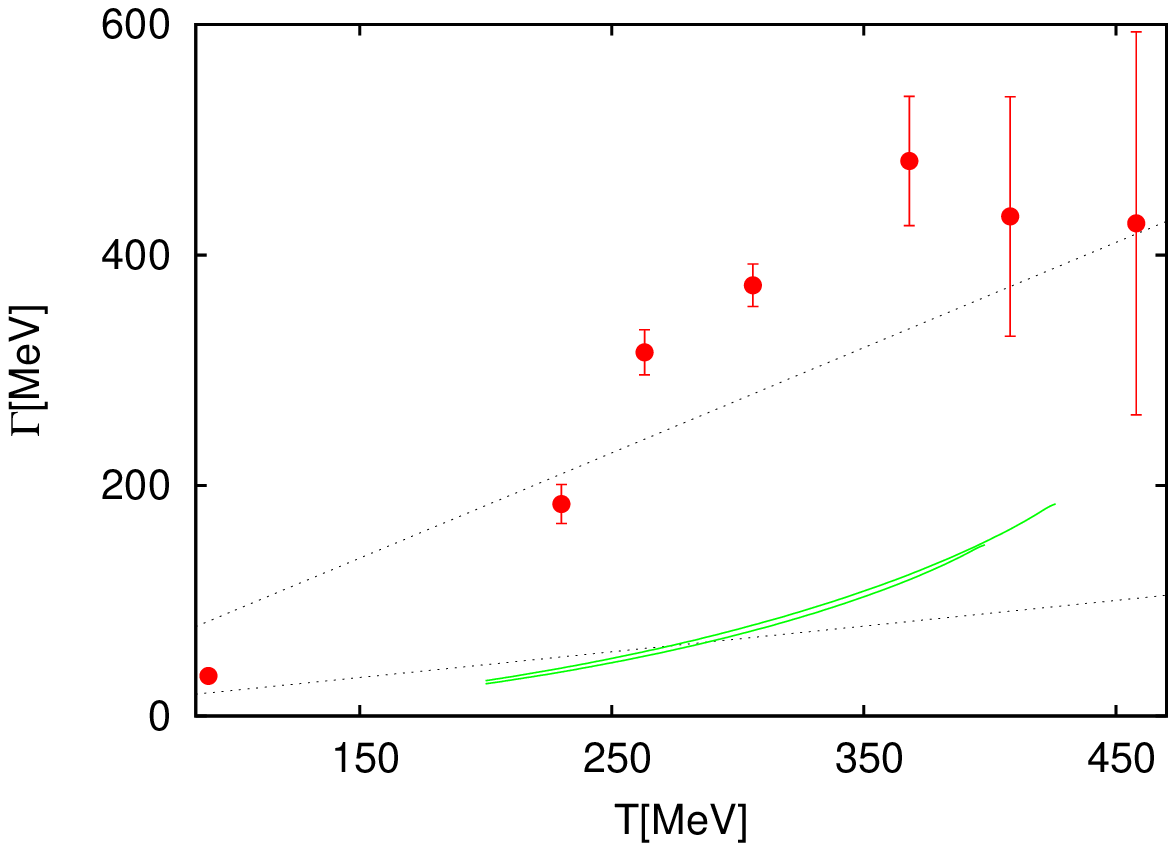}
\caption{(Left) Spectral function of the $\bar{b} \gamma_\mu b$ current
  at various temperatures, extracted from lattice correlators \cite{aarts1}. 
A strong peak for $\yna$ survives even at $\sim 2 \tc$. 
(Right) Width of the $\yna$ peak at various temperatures, extracted from 
lattice correlators \cite{aarts1}. Also shown are (green,solid line) results 
of a calculation from HTL potential (see Sec. \ref{sec.potential} 
and Fig. \ref{fig.impot}), and (black, dotted line) 
the trend from leading order of perturbation theory 
\cite{brambilla2,aarts1} for $\alpha_s$ = 0.25 (lower line) and 0.4 
(upper line).}
\label{fig.upsilon}
\end{center}
\end{figure}

In the bottomonia sector, there have been interesting recent studies 
\cite{aarts1,aarts2} using the formalism of non-relativistic QCD (NRQCD) on 
lattice \cite{lepage}. Use of NRQCD has two advantages in this context. 
In NRQCD the heavy quark mass $m_Q$ is much larger than all other
scales, and one replaces $\omega$ in Eq. (\ref{eq.kernel}) by $2 M_Q +
\omega$. Then for $M_Q \gg T$, one replaces the periodic kernel 
Eq. (\ref{eq.kernel}) by a simple exponential, $\exp(-\omega
\tau)$. Therefore independent correlator data is now available for the
whole range $[0, \beta)$. Also since one is now studying only
  excitations around $2 M_Q$, the low-$\omega$ diffusion peak
  structure is absent. Ref. \cite{aarts1} calculated the correlators
  in this formalism, and applied Bayesian analysis,
  Eq. (\ref{eq.bayes}), to extract $\som$. They found that 
1S bottomonia survive at least till temperatures of 2
$\tc$; see Fig. \ref{fig.upsilon}.
On the other hand, the 1P peaks were found to dissolve right after
$\tc$ \cite{aarts2}. These results are qualitatively in agreement with
earlier, preliminary studies of bottomonia within the relativistic framework 
\cite{bottom}. 

The peak position and the decay width of the 1S states have also been
extracted from the NRQCD studies of the Euclidean correlator in
Ref. \cite{aarts1}.  The decay width, $\Gamma$, calculated in
Ref. \cite{aarts1}, is also shown in Fig. \ref{fig.upsilon} (for their
system, $n_f=2$ with $m_q$ close to the strange quark mass, $\tc$ is
estimated to be $\sim$ 220 MeV). A near-linear increase of $\Gamma$
with temperature is seen. It is interesting to note that in effective
field theory calculations at weak coupling and $\al \mq \gg T$,
$\Gamma \sim 14 \: \alpha_s^3 \: T$ in leading order
\cite{brambilla2,aarts1}.  Within the rather large systematics, the
lattice data is roughly in agreement with this for $T \gtrsim$ 250
MeV, for $\alpha_s \sim 0.4$. Note that Fig. \ref{fig.upsilon}
indicates a large decay width for the 1S bottomonia already above
$\tc$. The origin of such a decay width would be collisions with the
thermal quarks and gluons in the medium.  Of course, the systematic
error associated with the extraction of width from the Euclidean
correlator is large, and Ref. \cite{aarts1} suggests that the
calculated widths should be treated as an indicative upper limit.
Preliminary results from another study, which also uses NRQCD
formalism to calculate bottomonia correlators, has reported much
smaller widths at comparable temperatures \cite{kim}.

While lattice NRQCD provides a very promising way to study bottomonia
in medium, it is fair to say that the studies are still reasonably 
recent and various systematics need to be better examined. 

\subsection{Nonrelativistic approach and ``potential at finite T''}
\label{sec.potential} 
A direct extraction of the spectral function from correlation
functions calculated in lattice QCD is the most direct approach to
understanding the behavior of quarkonia in deconfined plasma. 
Unfortunately, as discussed in the previous section, at the moment the
systematics of such a study are not in complete control. This is
likely to change with time. However, it will surely help to
supplement this direct approach with insights gained from other studies. 

A nonrelativistic potential approach has been remarkably successful in
quarkonia spectroscopy; it is therefore natural that several attempts
have been made to extend such an approach to finite temperatures.  The
first step towards this was the identification of the free energy cost
of putting an isolated $\qq$ pair in a thermal medium
\cite{macben}. As the distance between the $Q$ and the $\bar{Q}$
increases, in the confined medium the free energy cost of introducing
such a pair also increases. As $m_Q \to \infty$, the quark
is static and the effect of such a quark is approximated by a phase factor
$\sim e^{-m_Q \beta}$ multiplying a timelike gauge connection, called a
Polyakov loop, $L$, defined as
\begin{subequations}
\begin{eqnarray}
P(x) & = & \prod_{i=0}^{N_\tau-1} U_0(\vec{x}, i) \label{eq.upol} \\
L(x) & = & {\rm Tr} \: P(x). \label{eq.pol}
\end{eqnarray}
\label{eq.ploop} 
\end{subequations} 
After suitable mass renormalizations, one can define the free energy
cost mentioned above: 
\beq
\mathcal{F}^{\rm av}_{\qq}(\vec{x}) \ = \ \log \langle L(x) L^\dagger(0)
\rangle
\label{eq.avfree} \eeq
Note that $L(x)$ is invariant under color rotations, and therefore,
the free energy defined in Eq. (\ref{eq.avfree}) includes an averaging
over the color orientations of the $Q$ and $\bar{Q}$. If one wants to
study a $Q$ and $\bar{Q}$ in a singlet combination, as in 
quarkonium, it can be defined using $P(x)$, Eq. (\ref{eq.upol}) 
\cite{nadkarni}: 
\beq
\mathcal{F}^{\rm sing}_{\qq}(\vec{x}) \ = \ \log \langle Tr P(x) P^\dagger(0)
\rangle
\label{eq.avsing} \eeq
The right hand side in the above equation is not gauge invariant, and
therefore, needs to be calculated after fixing to a gauge. 
In perturbation theory, it can be shown that 
$\mathcal{F}^{\rm sing}_{\qq}(\vec{x})$ is gauge invariant, and in
leading order, has the expression
\beq
\mathcal{F}^{\rm sing}_{\qq}(\vec{x}) \ = \ - \frac{4}{3}
\frac{\alpha_s}{r} \: e^{-m_D r} .
\label{eq.pertf} \eeq
It has been evaluated also nonperturbatively, using
Eq. (\ref{eq.avsing}) in the Coulomb gauge \cite{zantow}.

$\mathcal{F}^{\rm sing}_{\qq}(\vec{x})$ has been widely used in the
literature to study the fate of quarkonia in QGP, by using it as a
finite temperature potential; it denotes, though, a free energy, and
not a potential. There has been considerable effort over the last
decade towards constructing a suitable potential for quarkonia at
finite temperature, in a field-theoretic framework
\cite{laine,pnrqcdt,blaizot}.
In vacuum, the effective field theoretic framework for formally
defining such a potential relies on the hierarchy of scales, $\mq ~
\gg ~ \mq v \sim 1/r ~ \gg E_b \sim \mq v^2 \sim g^2/r$, where $v
\ll 1$ is the relative velocity of the heavy quark, $Q$, and antiquark,
$\bar{Q}$, in the bound state, and $E_b$ is the binding energy.
The idea is to write
down an effective theory containing only the degrees of freedom
relevant for $\qq$ near threshold, i.e., those at scale $E_b$. 
So scales of $\mq$ and
$\mq v$ are integrated out. Integrating out of the scale $\mq$
leads to standard non-relativistic QCD (NRQCD), while further
integrating out $\mq v$ leads to so-called potential NRQCD (pNRQCD)
\cite{pnrqcd}, with Lagrangian \\
\begin{eqnarray}
\mathcal{L}_{\rm pNRQCD} &=& S^\dagger(\vec{r}) \left( i
\partial_0 - \frac{p^2}{2\mq} - V_S(r) ~ + {\rm corr.} \right)
S(\vec{r}) \nonumber \\ 
&+& O^\dagger(\vec{r}) \left( i
D_0 - \frac{p^2}{2\mq} - V_O(r) ~ + {\rm corr.} \right)
O(\vec{r}) + ...
\label{eq.pnrqcd} \end{eqnarray}
where $S, O$ denote the $\bar{Q} Q$ in singlet and octet
representations, respectively, and $V_{S,O}$ denote the corresponding
potentials. The $...$ include the singlet-octet transition terms. For
sufficiently heavy quarks such that $\mq v \gg \Lambda_{\scriptstyle
 \rm QCD}$, the parameters of $\mathcal{L}_{\rm pNRQCD}$ can be obtained 
perturbatively.
 
At finite temperatures, a new set of scales related to the temperature
$T$ are introduced. At very high temperatures, one can write down a
hierarchy $T ~ \gg ~ m_D \sim g T ~ \gg ~ g^2 T$, where $m_D$ is the
scale of screening of static charges and $g^2 T$ is the inherently
nonperturbative magnetic scale. Integrating out the scale $T$ leads to
the standard HTL (hard thermal loop) Lagrangian \cite{pisarski}. The
form of the finite temperature potentials depend on the relative
hierarchy of the thermal scales and the scales related to $\mq$
\cite{pnrqcdt}.

Let us take $\mq \gg T \gg \mq v$. Integrating out $\mq$ and $T$ then
leads to the HTL version of NRQCD. If $\mq v \sim m_D$, integrating out
these scales leads to a potential which was first derived in 
Ref. \cite{laine} slightly differently. The spectral function relevant to
the dilepton peak is connected to the fourier transform of the
real-time correlator \\
\beq
C^>(t,\vec{x}) = \int d^3x \langle J^\mu(t,\vec{x}) \:
J_\mu(0,\vec{0}) \rangle
\label{eq.cgt} \eeq
where $J^\mu(t,\vec{x})$ is the point vector current defined after
Eq. (\ref{eq.sigma}). Replacing $J_\mu$ in Eq. (\ref{eq.cgt}) 
by a point-split current 
\beq
J^{\rm split}_{\mu} (t,\vec{x}; \vec{r}) \ =
\ \bar{\psi} \left(t,\vec{x}+\frac{\vec{r}}{2} \right)  \: \gamma_\mu 
\: U \left(t;\vec{x}+\frac{\vec{r}}{2}, \vec{x}-\frac{\vec{r}}{2} \right)
\: \psi \left(t,\vec{x}-\frac{\vec{r}}{2} \right) 
\label{eq.split} \eeq
it is easy to check that in the non-interacting theory for non-relativistic
quarks, $C^{>}_{\rm split}(t,\vec{x}; \vec{r})$ satisfies a
Schr\"odinger-like equation \\
\beq
\left( i \partial_t - \left( 2m_Q - \frac{\nabla_r^2}{2 m_Q}
\right) \right) C^{>}_{\rm split}(t,\vec{x}; \vec{r}) \ = \ 0
\qquad ({\rm to} \ \mathcal{O}(1/m_Q^2)).
\label{eq.schr} \eeq
In the interacting theory, one can define a potential by equating the
left hand side of Eq. (\ref{eq.schr}) to $V(t, \vec{r}) C^{>}_{\rm
    split}(t,\vec{x}; \vec{r})$. Taking the static limit, $m_Q \to
\infty$, one gets
\beq
i \partial_t \: W(\vec{r}, t) \ = \ V(\vec{r}, t) \: W(\vec{r}, t).
\label{eq.lainepot} \eeq
where $W(\vec{r}, t)$ is the timelike Wilson loop. 
Going to the long time limit leads to
\begin{eqnarray}
V(r) \ & = & \  - \frac{4}{3} \alpha_s \: \left(\frac{e^{-m_D r}}{r}+
m_D \right) - i \frac{8}{3} \alpha_s T \: \Phi(r)  \label{eq.impot} \\
\Phi(r) \ & = & \int_0^\infty  \frac{dz \ z}{(z^2+1)^2} \ (1-\frac{\sin zr}{zr})
\nonumber
\end{eqnarray}
in leading order HTL approximation \cite{laine,pnrqcdt}.

\begin{figure}[ht]
\begin{center}
\includegraphics[width=0.4\textwidth,height=0.36\textwidth]{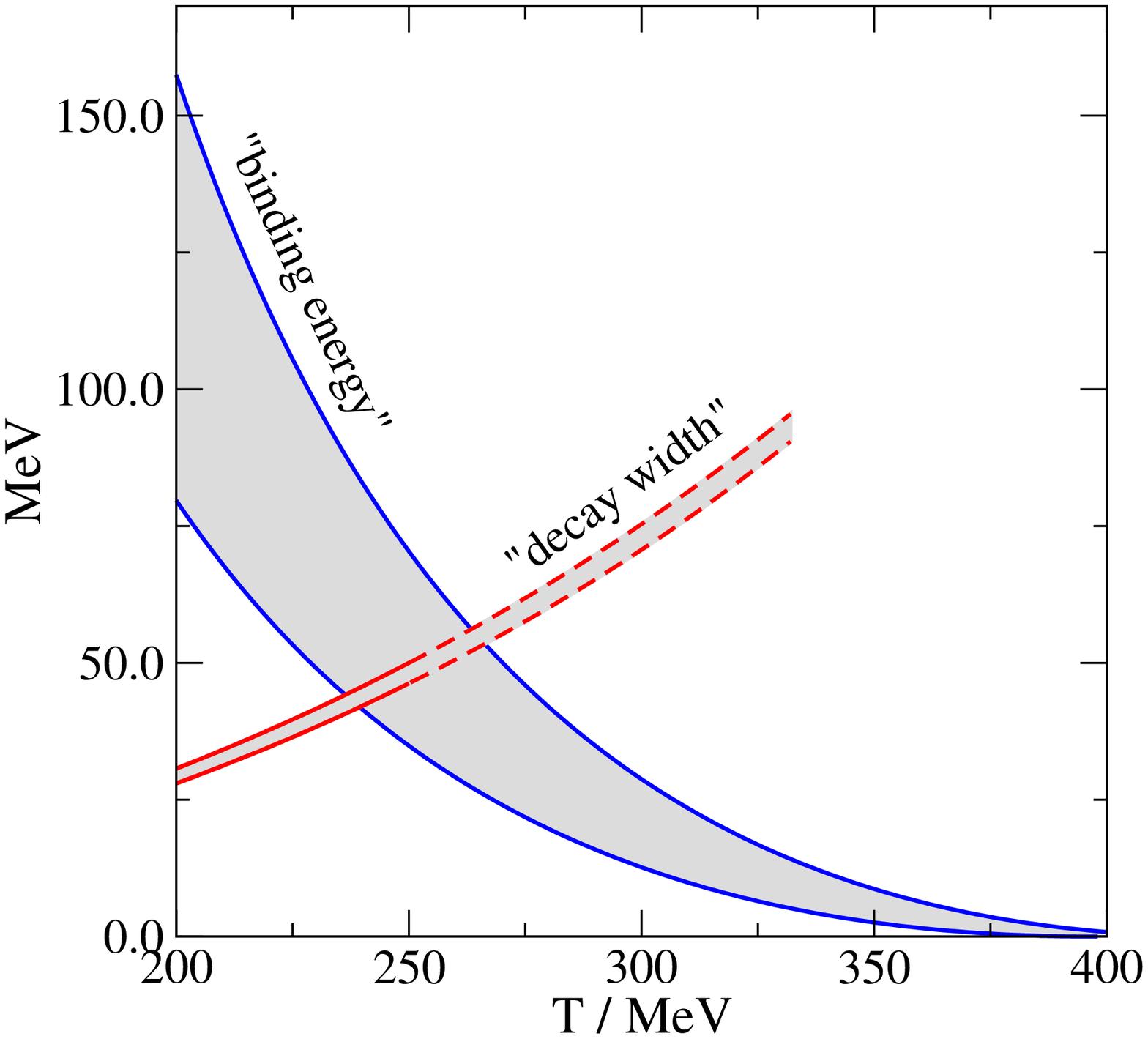}
\includegraphics[width=0.42\textwidth,height=0.38\textwidth]{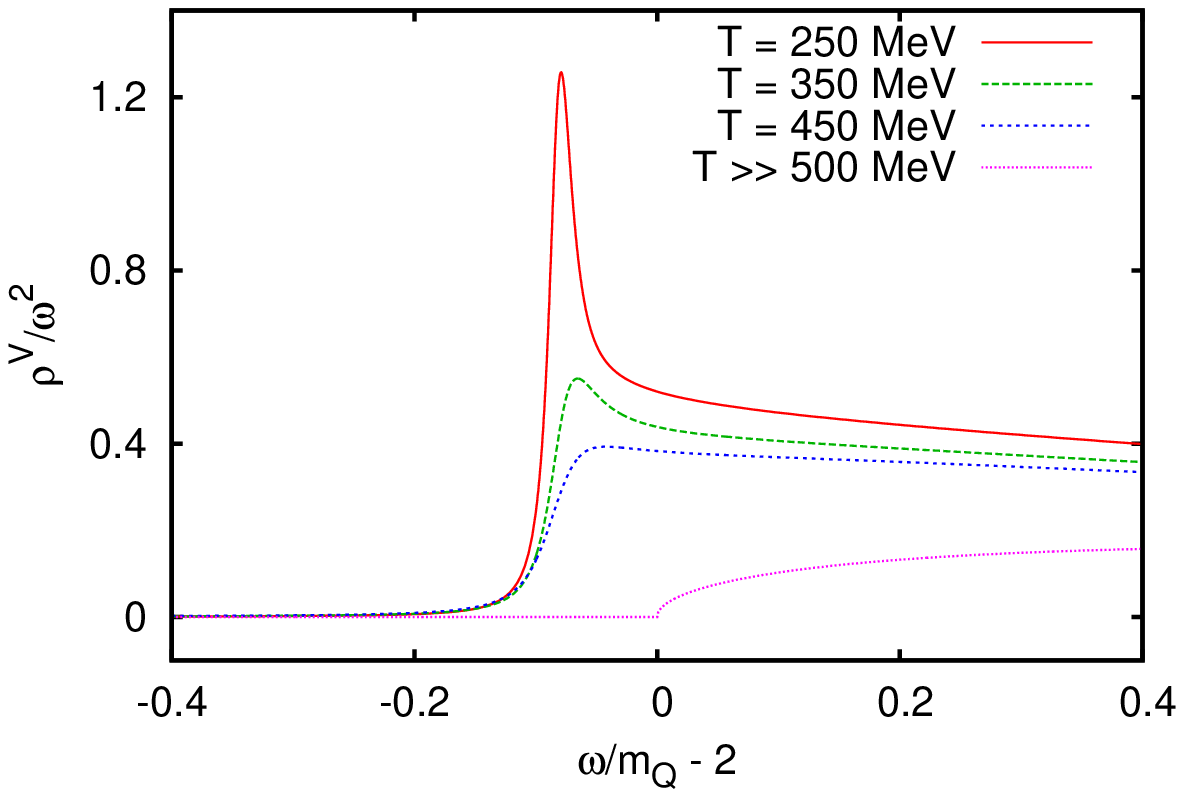}
\caption{(Left) Binding energy and decay width of the $\yn$ peak in 
the spectral function of the current $\bar{b} \gamma_i b$, using the
leading order HTL-resummed potential. From Ref. \cite{laine}. (Right)
Spectral function calculated directly from the same potential
\cite{vepsalainen}. Here $\mq$ = 4 GeV.}
\label{fig.impot}
\end{center}
\end{figure}

While for the plasma created in RHIC and LHC, the validity of the weak coupling
approximation used in reaching Eq. (\ref{eq.impot}) is questionable, it
is instructive to examine some features of the potential. 
The real part of $V(r)$ is, modulo a constant, identical to the
debye-screened singlet free energy  in
Eq. (\ref{eq.pertf}). Interestingly, there is also an
imaginary part to the potential. The imaginary part leads to a thermal
width $\propto \alpha_s T$ of the spectral function peak, which
incorporates the physics of Landau damping \cite{brambilla1}. To get
an idea of its contribution, Ref. \cite{laine} treated the imaginary
part as a perturbation, to get the decay width. For $\yn$ it was found 
that already at not-too-large temperatures $\sim$ 250 MeV, while the 
bound state survives, it acquires a considerable thermal width 
(see Fig. \ref{fig.impot}). This thermal
width, though, is smaller than that extracted in Ref. \cite{aarts1}
from lattice correlators (see Fig. \ref{fig.upsilon}). 

$V(r)$ (eq.\ref{eq.impot}) was also used to calculate the
spectral function, which is simply related to the Fourier transform of 
$C^>(\vec{r} \to 0, t)$ \cite{vepsalainen}.
It was found that for quarks in the bottom mass region, the
spectral function peak is severely depleted and broadened already by
$T \sim$ 350 MeV, and by a temperature of 450 MeV, no significant peak
is visible (see Fig. \ref{fig.impot}). If one uses the potential for
charmonium (where the separation of scales required is highly
questionable), one finds that already at $T \sim$ 250 MeV the peak
structure is essentially absent. A similar study was also carried out
in Ref. \cite{miao}, where also various systematics were studied. The
basic results are similar to Ref. \cite{vepsalainen}. Also the later
study highlighted the major role played by the imaginary part of the
potential in broadening the peak. Ref. \cite{miao} also calculated the
Euclidean correlators, and found that they do not completely agree
with the bottomonia correlators obtained from lattice.

As already mentioned, the hierarchy $T \gg \mq v$ is not valid at the 
temperatures of
interest in RHIC and LHC. At least for the $\Upsilon$, one in fact expects
$\mq v \gg T$. In such a case, one first integrates out the scale $\mq
v$ from NRQCD, to get the
pNRQCD action, Eq. (\ref{eq.pnrqcd}). Further integrating out the scale 
$T$ then leads to thermal corrections which are very different from
Eq. (\ref{eq.impot}) \cite{pnrqcdt}. If $m_D \gg E_b$, the thermal 
potential is still well-defined, but the real part of the potential
does not have the screened form. The potential still gets an imaginary
component, which now has two main components: a term $\propto \al^3 T$
which comes from a transition to color-octet state, and terms like 
$\al T m_D^2 r^2$ which are related to Landau damping. On the other
hand, if $E_b \gg m_D$, the thermal potential is not well-defined. Of
course, thermal corrections to the binding energy and the the decay width are
well-defined quantities, and have been calculated in weak coupling
\cite{brambilla2}. The thermal decay width once again has
contributions $\propto \al^3 T$ related to singlet-to-octet
transition, and terms $\propto \al T m_D^2 r_0^2$ related to Landau
damping \cite{brambilla2}, where $r_0 = 3/(2 \al \mq)$ is the Bohr radius.

One message to take, both from the effective field theory studies and
from studies of the previous section, is that even before the dissolution
of the quarkonia peak, the states can get a substantial thermal decay
width.  Rather than a single ``dissolution temperature'', it is the
temperature-dependent width which is of phenomenological significance.
The decay width obtained from the imaginary part of the potential has
been compared to phenomenological estimates of quarkonia dissociation
in Ref. \cite{brambilla1}. 

In the discussion so far, we have stressed that the imaginary part of
the ``potential'' is a theoretical tool to describe the broadening
of the quarkonium structure in the spectral function,
due to interactions with the thermal gluons and quarks. Its
interpretation has been further clarified using the language of open
quantum systems \cite{borghini,akamatsu}. Starting from a complete
description of the $\qq$ + thermal medium, one can integrate out the
thermal medium to get an effective description of the $\qq$ system in
medium. Integrating out of the medium leads to noise terms in
the description of the $\qq$ system, which cause both
dissipation of the heavy quark, and a lack of coherence between the
$\qq$ pair in quarkonia, leading to an increased thermal width. 
This has been worked out in perturbation theory, to give the
same complex potential as above \cite{akamatsu}.    
 
One can also try to calculate the potential, Eq. (\ref{eq.lainepot}),
nonperturbatively, without assuming weak coupling or any particular
ordering between the scales $T$ and $\mq v$.  Ref. \cite{burnier}
calculated the Euclidean timelike Wilson loop, $W(\vec{r}, \tau)$, and
then employed an analytic continuation similar to that described in
Sec. \ref{sec.lattice} to obtain the potential from it. A Bayesian
analysis similar to, but not identical to, MEM was used
\cite{rothkopf}, which probably needs careful examination by others. 
A complex potential was extracted from the Euclidean Wilson loop
calculated in a gluon plasma. At 2.33 $\tc$, both the real and the
imaginary parts of the potential extracted from the data are
considerably different from the hard thermal loop calculation. Note
that the timelike Wilson loop is identical to the Polyakov loop
correlator in Eq. (\ref{eq.avsing}), calculated in the axial
gauge. Ref. \cite{burnier} also calculated the Polyakov loop
correlator in Coulomb gauge. Analyzing it in the same way, they
obtained a potential which is much closer to the perturbative
calculation of Ref. \cite{laine}. Note however that the coulomb gauge
calculation was done in great detail already \cite{zantow}; it is not
clear how sensitive the correlation function is to the imaginary part
of the potential.

\section{Prediction for quarkonia production in relativistic 
heavy ion collision}
\label{sec.hic}

In the previous section, we discussed various calculations that
investigate the fate of a quarkonium put as a probe in static,
equilibriated QGP. Of course, the experimental situation is very
different. A $\qqc$ pair gets formed, probably in a hard collision at
early times; the quarkonium state gets formed, either in the
pre-equilibrium stage or in the plasma. Also the system is not static,
the temperature profile changes. While understanding the behavior of
an external $\jpsi$ in static plasma forms the first step to
predicting the $\jpsi$ production in relativistic collisions, one
needs to put it in the context of the fireball created in a heavy ion
collision.  In this section we discuss some calculations towards
quantitative prediction for $\jpsi$ production in heavy ion collision,
and describe some ingredients for such a calculation.

\subsection{``Regeneration'' and thermal quarkonia}
\label{sec.reg}
The spirit of the discussion of Sec. \ref{sec.theory} was that the
$\jpsi$ is formed very early in the plasma, in a way possibly
similar to that in a $pp$ collision, and we investigate the survival
probability of the $\jpsi$ in the QGP. The hadrons made of light
quarks, on the other hand, are described very well by the assumption that
as the medium cools to below deconfinement, the quarks present in the
medium coalesce to form hadrons according to a thermal distribution. If the
density of $c, \bar{c}$ is sufficiently high in the medium, 
one also needs to investigate the
possibility that at freezeout a $c$ and a $\bar{c}$ coalesce to form a
$\jpsi$ \cite{reg}. In the literature, $\jpsi$ production this way is
dubbed ``regeneration''.

In regeneration calculations \cite{reg,reg2}, the different hadrons with 
(open or hidden) charm are distributed statistically, just as the 
light hadrons are. The charm quarks are produced as $\bar{c} c$ pairs 
in initial hard collisions, but then will develop in the plasma as colored
$c(\bar{c})$ quarks. At the time of freezeout, they then hadronize 
according to a statistical thermal distribution. So the ratios of,
e.g., different charmonia will follow a statistical
distribution. An early motivation was the fact that the ratios of the 
production cross-section of the $\psib$ and the $\jpsi$ in the 158A
GeV Pb-Pb collisions in SPS followed a statistical
distribution \cite{reg}.  

The number of $\jpsi$ produced will be given as 
\begin{eqnarray}
N_\jsub \ & = & \ g_c^2 \: V \: n^{\rm th}_{\jsub}(T_{\rm fr},
\mu_B), \label{eq.reg} \\
{\rm where} \ \ n^{\rm th}_i &=& g_i \: \int \frac{d^3p}
{e^{(E_i(p)-\mu_i)/T} \: \mp \: 1} \label{eq.therm}
\end{eqnarray}
is the thermal (Bose/Fermi)
distribution function for hadron $i$, $T_{\rm fr}$ and $\mu_B$ are the
chemical freezeout temperature and baryon chemical potential, V is the
fireball volume, and $g_c$ is an extra degeneracy factor to take into
account the fact that the production of the charm quarks happen in
hard collisions and the total number of charm quarks is much larger
than it would have been if the charm quarks were chemically
equilibriated in the plasma \cite{reg}. 

The model is simple, and with clear predictions for rapidity and $p_T$
distributions of the $\jpsi$. The earlier works on this model
\cite{reg} found
good agreement of the relative yields of $\jpsi$ and $\psi^\prime$ in
the highest energy Pb-Pb runs of SPS. More recent studies
\cite{reg2} have also found good agreement of this ratio in RHIC. 
The early studies \cite{reg} had also predicted little 
or no suppression of $\jpsi$ production in RHIC, and substantial
enhancement in LHC, compared to scaled $pp$ collision results. This 
has not been borne out by data. More detailed recent
studies \cite{reg2}, which take into account the possibility that
even in collisions that lead to a plasma in the core, there may be 
hard collisions between (surface) nuclei which do not lead to plasma
formation, have reported good agreement with RHIC data for $\jpsi$
production.  

A generic prediction of the regeneration calculations is more
suppression in the forward rapidity region than in midrapidity. This
is in qualitative agreement with the trend seen in 200A GeV Au-Au
collisions in RHIC.  Also, as Eq. (\ref{eq.therm}) suggests, the $p_T$
distribution will be rather steep. Assuming same $T_{\rm fr}$ as light
hadrons, a much steeper $p_T$ distribution is predicted than seen in
the experiment \cite{reg2}. A much larger freezeout temperature $\sim$
360 MeV, compared to the initial temperature, was required in
Ref. \cite{reg2} to describe the $p_T$ dependence in RHIC.

A variant of the regeneration mechanism has been suggested in
Ref. \cite{shuryak}. In their picture, while the initially produced
$\qqc$ pair does not lead to a $\jpsi$ formation in the plasma, they do
not become completely decorrelated. The motion of a heavy quark in
plasma can be understood as a diffusion process
\cite{moore,diffusion}, with a rather small diffusion coefficient
\cite{diffusion,francis}. The smallness of the diffusion coefficient,
combined with the interaction between the $\qqc$ pair, lead to the
$c$ and $\bar{c}$ staying spatially correlated through their evolution 
in the fireball. This, the authors argue, leads to a $\jpsi$
cross-section larger than a naive regeneration calculation suggests, and a
$p_T$ distribution similar to that of the original $\qqc$ pair.  

Some authors (e.g., \cite{grandchamp}) have used a combination of
directly produced and recombined $\jpsi$ to explain the $\jpsi$ yield,
e.g., in RHIC. In \cite{grandchamp}, the $\jpsi$ yield in SPS is
almost completely directly produced. In RHIC, because of the hotter
and larger fireball, a larger fraction of the directly produced
$\jpsi$ is suppressed, but some $\jpsi$ are regenerated,
giving a final suppression factor similar to SPS. 

Another intuitive signature of regeneration would be elliptic flow of 
$\jpsi$ \cite{nu}. The charm quark shows substantial elliptic flow in RHIC
\cite{diffusion}. If the $\jpsi$ is regenerated, it is expected to inherit 
this elliptic flow. On the other hand, a color singlet $\jpsi$ moving through
the plasma will not show a substantial flow. While no significant elliptic 
flow of $\jpsi$ was seen in RHIC, elliptic flow has been measured in LHC
\cite{aliceflow} and has been used to argue for substantial regeneration 
of $\jpsi$ in the fireball created in LHC \cite{zer}.

Due to the large mass of the $b$, regeneration is usually considered
to play a small role in the bottomonia sector. It has, however, been
pointed out that the ratio of the different $\Upsilon(nS)$ states can
be explained by regeneration, assuming a $T_{\rm fr} \sim$ 250 MeV \cite{gupta}.

\subsection{Quarkonia in the fireball}
\label{sec.actual}
To calculate quarkonia production cross-section in the fireball
produced in the relativistic heavy ion collisions, e.g., the $\jpsi$
or $\yn$ peak in dilepton channel, we need to incorporate the inputs
from the previous sections in the framework of the evolution of the
fireball. For use of quarkonia as a marker of deconfinement, as
originally envisaged by Matsui and Satz \cite{matsui}, one needs to 
have a theoretical calculation of quarkonia production for a given initial
temperature of the plasma. For such a quantitative prediction, 
we need to have an understanding of the following processes. 

\begin{itemize}
\item
The production of $\bar{c} c$ pair, possibly in a hard $g g$
collision. 
\item
Connecting the $\bar{c} c$ pair to the $\jpsi$ resonance.
\item
Fate of the $\jpsi$ as it moves in the plasma, which is expanding
and cooling.
\item
Fate of other $\qqc$ resonances which can decay to $\jpsi$.
\item
Possibility of generation of $\jpsi$ at the freezeout.
\end{itemize}

Usualy the quarkonia yield in $A$-$A$ collisions is presented as $R_{AA}$,
ratio of the yield in $A$-$A$ collision with the scaled yield in $pp$
collision (for the same window of variables like $p_T, y$ etc): \\
\begin{equation}
R_{AA} \, (\jpsi) \ = \ \frac{N_{AA}(\jpsi)}{N_{\rm coll} \:
  N_{pp}(\jpsi)}
\label{eq.raa} \end{equation}
where $N_{\rm coll}$ is the number of binary collisions. A deviation
of $R_{AA}$ from 1 does not necessarily indicate medium effect.
The production of the $\bar{c} c$ is a hard process; but the
gluon distribution function is a nonperturbative input. These
distribution functions can be different in the nucleus from that in
the proton; e.g., the low $x$ rise of the gluon distribution function
can be tempered due to two low $x$ gluons fusing
(``shadowing''). Usually one extracts the distribution functions in
nucleus from inputs like deep inelastic $e$-$A$ collisions as well as
observables like dilepton and pion production in $p$-$A$ collisions
\cite{eps09}. 

The conversion of the $\bar{c} c$ to $\jpsi$ is a complicated process
even in the vacuum \cite{qwg}. Many calculations use the simple
``color evaporation model'', where production cross-section of the
$\jpsi$ to the $\qqc$ production cross section is given simply as
\beq 
\sigma_{\jsub}(s) \approx g_{\qqc \to \jsub} \: \sigma_{\qqc} (s)
\label{eq.cem} \eeq
with $g_{\qqc \to \jpsi}$ energy independent \cite{cem}. Similar
relations are written for the other charmonia. A more rigorous aproach
is to use nonrelativistic QCD (NRQCD) \cite{bodwin}. One uses a
separation of the scales $m_Q$ and $m_Q v$ to write down the $\jpsi$
as a superposition of a singlet $\qqc$ state and states where $\qqc$
are in an octet configuration, combining with $g$ to form a color
singlet. The original $\bar{c}c$ can form in either color singlet or
color octet, and then evolves into the $\jpsi$ by emitting gluons. If
we estimate the formation time of the $\jpsi$ as $\tau_{\scriptstyle
  \jpsi} \sim 1/E_b$, the binding energy of the $\bar{c} c$ pair, we
get $\tau_{\scriptstyle \jpsi} \sim$ 0.5 fm, which is of the order of
the formation time of the plasma. For $\jpsi$ with large $p_T$, time
dilation increases the formation time further. So the in-medium
behavior of the precursor to $\jpsi$ needs to be understood
\cite{kharzeev,rishi}. In particular, for large $p_T$ $\jpsi$ the
precursor is mostly in color octet state, and it has been argued that it
interacts much more readily with the medium, leading to dissolution
\cite{kharzeev,reg2} or quenching of $p_T$ \cite{rishi}.

The interaction of the $\jpsi$ (and other quarkonia, which may decay
into $\jpsi$) with the medium is
probably the most studied part of the scheme outlined above. Following
the original intuitive argument of Matsui and Satz \cite{matsui}, many
early works used a dissociation temperature, usually from using the
singlet free energy in the Schr\"odinger equation. In a series of
papers, Kharzeev and Satz studied the dissociation of the $\jpsi$ (and
its precursor, the color octet state) through gluon dissociation,
generalizing the multipole analysis \cite{peskin} for thermal
gluons. For thermal gluons a free gluon gas distribution has been
used, which is probably not a good approximation at temperatures of
interest to RHIC and LHC. A similar approach has been followed in
Ref. \cite{rishi}.

Ideally, the temperature modification of $\jpsi$ and $\yn$ should be
incorporated by putting in the corresponding spectral function. That
is, however, more difficult; what has been done in
Refs. \cite{strickland,margotta} is to use the imaginary part of the potential
to calculate a thermal decay width, and evolve that through the
history of the plasma to calculate a suppression factor $R_{AA}$. On
the other hand, in Ref. \cite{rapp} the decay width is obtained from
the imaginary part of the quark propagator, which incorporates the
scattering T matrix, which can be evaluated self-consistently through
a Bethe-saltpeter equation \cite{rapp2}. The $\qq$ potential is an
input in the Bethe-Saltpeter equation.

It is worth mentioning here that in many studies, the real part of the
potential is replaced by an ``internal energy'' \cite{zantow2}, which
is obtained by subtracting an entropy term from the singlet free
enrgy, Eq. (\ref{eq.avsing}). This has largely been motivated by the
fact that use of the free energy seems to give too large a suppression
for both $\jpsi$ and $\yna$. As the discussion in
Sec. \ref{sec.potential} shows, however, there is little theoretical
justification for using such an ``internal energy'' for study of
quarkonia dissociation in plasma.

In order to quantitatively study the $\jpsi$ peak, it is not enough to
study the modification of the $\jpsi$ in the plasma. Almost half of
the $\jpsi$ seen in a $pp$ collision come from a ``feeddown'' route:
the original $\qqc$ pair goes to $\chi$ and $\psi(2s)$ states, which
have a substantial branching fraction to $\jpsi$.  
Some $\jpsi$ also come from decays of the $B$ mesons: at Tevatron,
this fraction has been estimated as $\sim 9 \pm 1 \% $ \cite{cdf}.  
The time scale for the $B \to \jpsi$ decay is
$ps$, and the $\jpsi$ coming from $B$ can be subtracted out; the
$\jpsi$ yield after such a subtraction is referred to as 
``prompt'' $\jpsi$ \cite{cdf,cms}. Time scales for the feeddown 
decays from $\chi_c$ and $\psib$ are $\sim 200$ fm or more. So these
states are expected to move through the medium as the excited
states. The fraction of $\jpsi$ coming from $\psib$ and $\chi_c$
states have been estimated as $\sim 9 \pm 3 \%$ and $\sim 30 \pm 7 \%$
at Tevatron, and similar values at lower energies \cite{faccioli}.
For the $\yna$, the CDF collaboration
has measured the feeddown fraction in Tevatron \cite{cdf2}: for $p_T >$ 8
GeV, the fraction of directly produced $\yna$ is about $51 \pm 12 \%$,
while about $11 \pm 8 \%$ come from decay of excited $\yn$ and $38 \pm
9 \% $ come from $\chi_b$ decays. Since the excited $\chi$ or
$\psib$ states are more readily dissolved by the medium
\cite{kms,digal,kharzeev,prd}, about
50\% suppression of the $\jpsi$ and $\Upsilon(1S)$ yield can come
simply from the melting of the excited states in medium into open charm.

\section{Experimental results}
\label{sec.expt}
There has been an immense body of experimental results on $\jpsi$ and
$\yn$, starting from the early experimental efforts to create
quark-gluon plasma. For completeness we mention some trends from the
experiments; detailed survey of experimental results can be found
elsewhere \cite{rossi,tserruya}.
 
$\jpsi$ suppression compared to $pp$ collisions was already seen at
the O-Cu and S-U collisions in the NA38/NA50 experiments in SPS,
CERN. However, this suppression could be completely understood in
terms of cold nuclear matter effect, like shadowing
(Sec. \ref{sec.actual}) and interaction of the $\jpsi$ with nuclear
matter, taking a nuclear absorption cross-section $\sigma^{\rm
  abs}_{\jpsi N} \approx$ 4 mb and $\sigma^{\rm abs}_{\psib N}
\approx$ 7 mb \cite{na50}.  The investigation of cold nuclear matter
effects can be done by conducting $pA$ collisions at the same energy.
This has now become a staple of the experimental program, and
understanding the quarkonia yield in such collisions is vital before
one can interpret the suppression in $AA$ collision.

\begin{figure}[ht]
\begin{center}
\includegraphics[width=0.45\textwidth,height=0.35\textwidth]{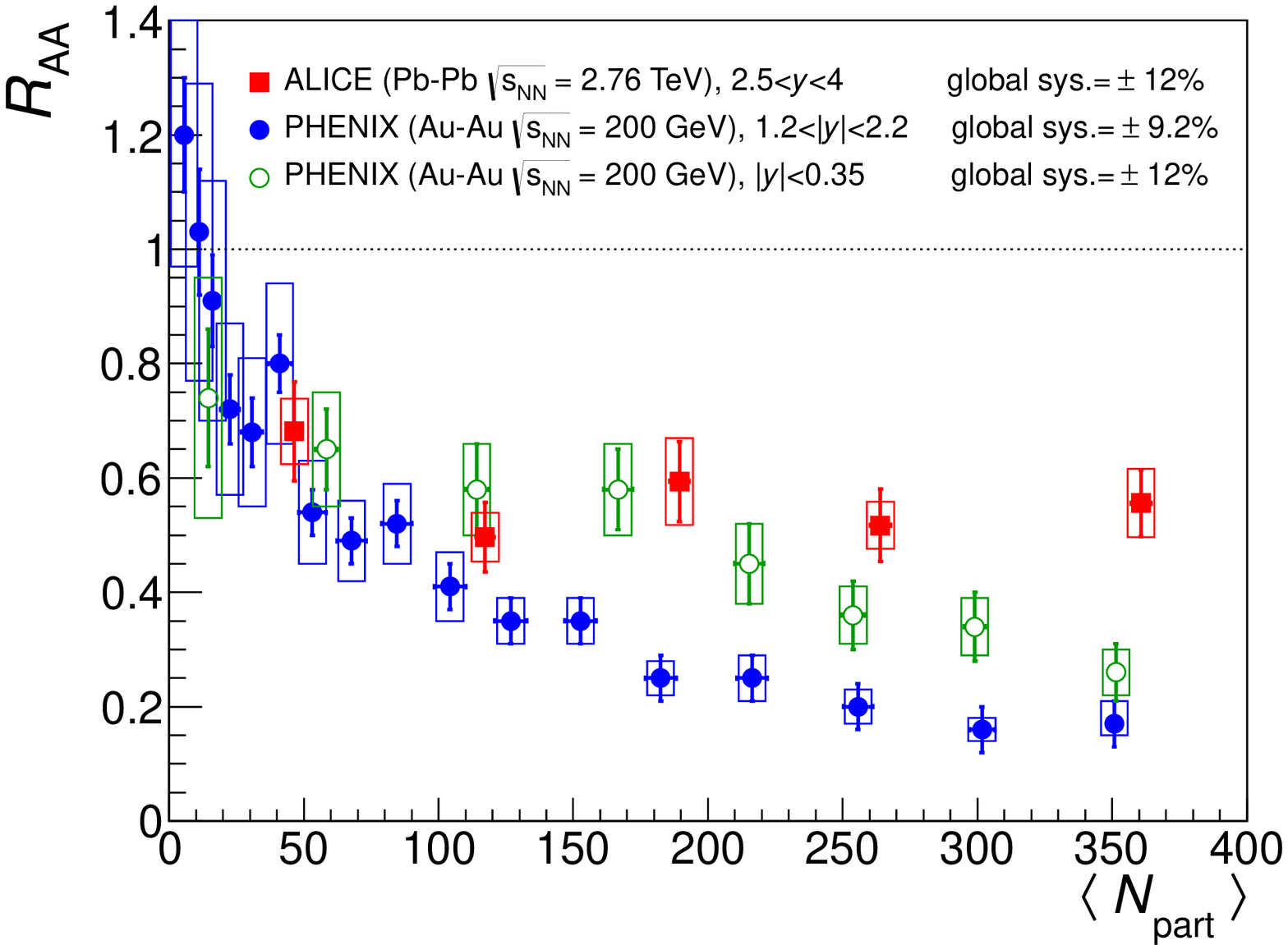}
\includegraphics[width=0.45\textwidth,height=0.35\textwidth]{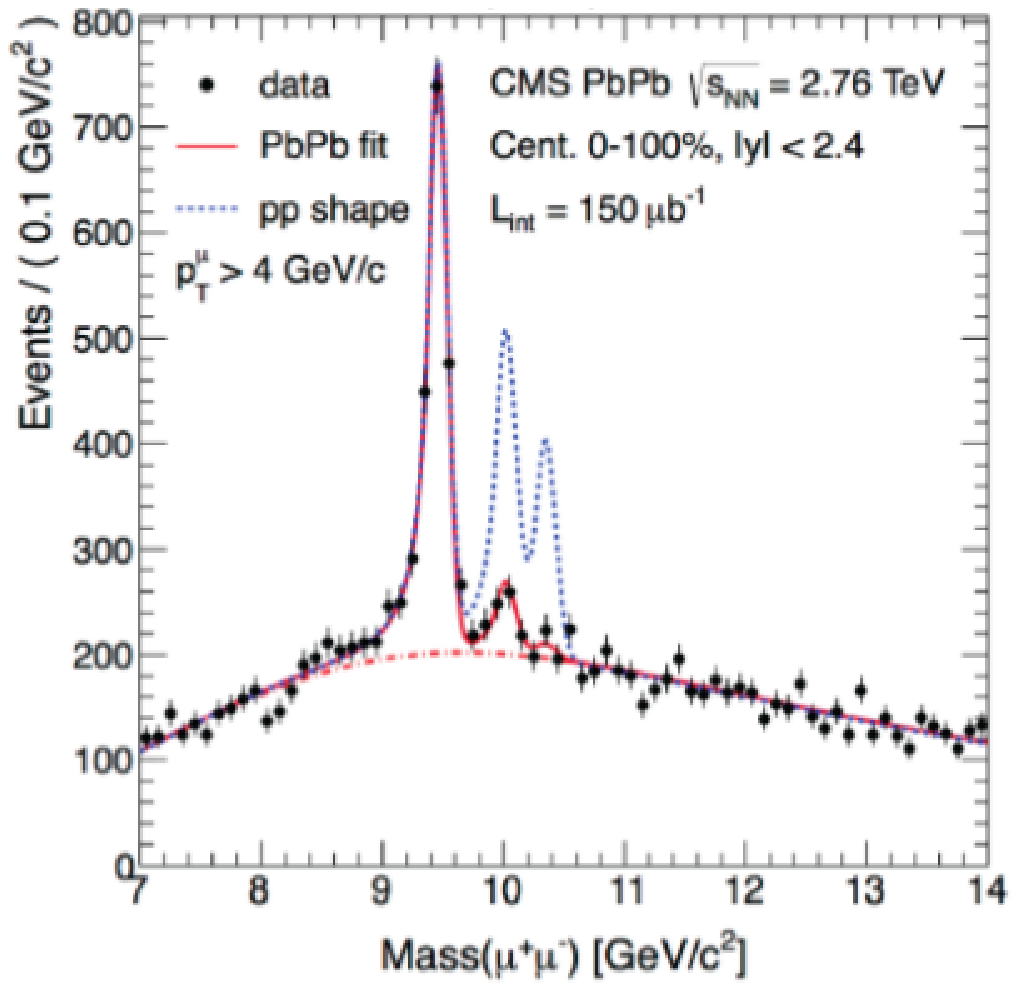}
\caption{(Left) $\jpsi$ $R_{AA}$ measured by the Alice experiment in
  the Pb-Pb collisions in LHC at 2.76 A TeV, compared to 200 A GeV Au-Au
  results from Phenix (\cite{alice}, \textcopyright American Physical Society). 
  (Right) $\yn$ peaks in the dimuon channel at midrapidity, from CMS 
\cite{tserruya,cmsY}.  The $\yna$ peak has been normalized to the $pp$
collision, showing the suppression of the $\ynb$ and $\ync$ states.}
\label{fig.exp}
\end{center}
\end{figure}

A larger suppression of the $\jpsi$ than what could be explained by cold
nuclear matter effects was observed by the NA50 experiment in 158 A
GeV Pb-Pb collisions in SPS. This suggested an onset of deconfinement
\cite{na502}. A large suppression has also been seen in the 200 A GeV
Au-Au collisions in RHIC (Fig. \ref{fig.exp}).  The level of
suppression seen was somewhat similar to that seen in SPS, which was a
surprise, given the much larger center-of-mass energy and the
expectation of a much longer living plasma.  One way to explain the
data was to assume that in both experiments the plasma was hot enough
to dissolve the excited $\chi_c$ and $\psib$ states, but not hot
enough to melt the directly produced $\jpsi$ \cite{kks}. 
The data could also be explained in other ways: e.g., it
is possible that a larger part of the directly produced $\jpsi$
was dissolved in the RHIC experiment, but there was some $\jpsi$
produced via recombination (Sec. \ref{sec.reg}), to keep the total
yield similar \cite{zhao}. Other attempts to explain the total yield
have used only regeneration \cite{reg2}. As explained in Sec. \ref{sec.reg},
regeneration calculations have been fairly successful in describing
the rapidity dependence of the $\jpsi$ suppression, but have had
difficulty explaining the $p_T$ dependence \cite{phenix,phenix2}.

The data from the Pb-Pb collisions at much larger center-of-mass
energy 2.76 A TeV (but in the forward rapidity region), as measured by
the Alice collaboration \cite{alice}, is also shown in
Fig. \ref{fig.exp}.  The data does not show a much stronger
suppression.  Data at midrapidity, but larger $p_T$, from CMS
\cite{cms} shows suppression at levels similar to the Phenix data in
Fig. \ref{fig.exp}.  A combination of suppression and recombination
have been suggested to explain the data \cite{zhao2}. However, details
of the $p$-Pb data for charmonia have not been completely understood
\cite{manceau,rossi}.

An interesting suggestion has been to look at not the $R_{AA}$ but the
ratio of $\jpsi$ with open charm cross-section \cite{sridhar,open}.
This will remove the uncertainties due to the nuclear distribution
functions.  The double ratio \cite{open} \\
\begin{equation}
S_{\jpsi} \ = \ \frac{g^{\scriptstyle AA}_{\qqc \to
    \jpsi}}{g^{\scriptstyle pp}_{\qqc \to \jpsi}}, \qquad \qquad
g_{\qqc \to \jpsi} \ = \ \frac{N(\jpsi)}{N(\qqc)}
\label{eqn.open} \end{equation}
then shows the medium modification of $\jpsi$ binding. Using this
quantity, Satz has claimed that the forward rapidity / large $p_T$
data of LHC does not show any anomalous suppression of $\jpsi$, while
the RHIC data in Fig. \ref{fig.exp} does \cite{open}.

A very beautiful measurement of the $\yn$ production from the CMS
experiment avoids many of the experimental uncertainties, and offers a
way of studying $\yn$ suppression in the Pb-Pb collisions in LHC. The
right panel in Fig. \ref{fig.exp} shows the $\yn$ peaks in the dimuon
channel, where the $\yna$ peak has been normalized to agree with the
peak in $pp$ collisions. Then the peaks for the $\ynb$ and $\ync$ are
considerably suppressed. In fact, the $R_{AA}$ value for $\yna$ is
quoted as $0.56 \pm 0.08 \pm 0.07$ \cite{cmsY}.  From the discussion
at the end of Sec. \ref{sec.actual}, this is consistent with no
suppression of the direct $\yna$ but almost complete suppression of
the feeddown component. This is in line with what one would expect if,
following the lattice studies, one expects the $\yna$ not to be
modified much in the plasma at moderate temperatures, while the
excited states melt readily (Sec. \ref{sec.lattice}). Note, however,
that the large decay widths shown in Fig. \ref{fig.upsilon} would
suggest a substantial suppression of the $\yna$ also.  The $R_{AA}$
for $\ynb$ is $0.12 \pm 0.04 \pm 0.02$ and that for $\ync$ is $< 0.1$
\cite{cmsY}, indicating major dissociation of these states.

\section{Summary and outlook}
\label{sec.summary}
In this article, the current status of our understanding of quarkonia 
yield in relativistic heavy-ion collisions is presented.
Conceptually, a lot of insight has been gained in the last
decade, and the simple picture of quarkonia dissociation due to debye
screening has been replaced by a detailed understanding of the 
dissociation mechanism from QCD.

Lattice QCD (Sec. \ref{sec.lattice}) has emerged as a preeminent tool
for understanding the behavior of quarkonia in static plasma. But
getting quantitative information about the spectral function, decay
width etc. have so far been difficult. While it is likely that
eventually we will be able to extract the physics from lattice
correlators, it probably will require some new ideas. One recent idea
has been non-relativistic QCD on lattice, which is a very promising
tool for bottomonia at least. Simultaneously, other approaches,
largely based on perturbative NRQCD, have clarified many
misconceptions (Sec. \ref{sec.potential}).  One development has been a
theoretically justified construction of a finite temperature effective
potential, and illustration of how it captures effects like thermal
gluon dissociation.
 
The calculation of quarkonia yield in the expanding fireball produced
in heavy ion collisions is more challenging. In Sec. \ref{sec.actual}
necessary steps are discussed, and some calculations
that try to incorporate many of the formal developments of 
Sec. \ref{sec.lattice} in them are outlined. Of course, for such a 
calculation one
needs to know the behavior of a quarkonium moving with respect to the
medium. Some preliminary calculations exist in that direction
\cite{momentum}, but clearly more needs to be done. Some other
major uncertainties pertain to the formation of the $\jpsi$, and
interaction of the medium with the precursor to the $\jpsi$. 
 
Almost three decades after the suggestion of quarkonia as a probe of
the deconfined medium, it remains a topic of great interest. While
still not a thermometer of the plasma as originally envisaged by Satz
and others \cite{kms}, it has been an invaluable source of insight
into the nature of the deconfined medium.

\acknowledgments
I would like to thank Mikko Laine and Jon-Ivar Skullerud for providing
me with data related to Figs. \ref{fig.upsilon}, \ref{fig.impot}.
The first draft of this article was completed during Whepp-13. I would 
like to acknowledge discussions with the participants of the meeting, in
particular with D. Das, S. Gupta, R. Gavai, R. Sharma, P. Shukla and
K. Sridhar.

\bibliographystyle{pramana}
\bibliography{references}

\end{document}